\documentclass[doublecol]{epl2}

\title{Interfacial Instability of Charged End-Group Polymer Brushes}
\shorttitle{Interfacial Instability of Charged End-Group Polymer Brushes}
\author{Yoav Tsori \inst{1} \and David Andelman\inst{2} \and Jean-Fran\c{c}ois
Joanny\inst{3}}
\shortauthor{Y. Tsori \etal}

\institute{ \inst{1} Department of Chemical Engineering, Ben-Gurion University of the Negev, P.O. Box 653, 84105 Beer-Sheva, Israel. \\
\inst{2} Raymond and Beverly Sackler School of Physics and Astronomy, Tel Aviv University,  Ramat Aviv, Tel Aviv 69978, Israel.\\
\inst{3} Institut Curie, UMR 168, 26 rue d'Ulm, F-75248, Paris Cedex 05, France.}
\pacs{61.25.Hq}{Macromolecular and polymer solutions; polymer melts; swelling}
\pacs{41.20.Cv}{Electrostatics; Poisson and Laplace equations, boundary-value problems}
\date{21/1/2008}

\newcommand{\eps}{\varepsilon}

\abstract{
We consider a polymer brush grafted to a surface (acting as an electrode) and bearing a charged group at its free end. Using a second distant electrode, the brush is subject to a constant electric field. Based on a coarse-grained continuum model, we calculate the average brush height and find that the brush can stretch or compress depending on the applied field and charge end-group. We further look at an undulation mode of the flat polymer brush and find that the electrostatic energy scales linearly with the undulation wavenumber, $q$. Competition with surface tension, scaling as $q^2$, tends to stabilize a lateral $q$-mode of the polymer brush with a well-defined wavelength. This wavelength depends on the brush height, surface separation, and several system parameters.}

\begin{document}
\maketitle


\section{Introduction}

There are different ways to bind polymers to surfaces. Either by adsorption from solution or grafting them onto the surface with a terminal group or having an adhering block in case of block copolymers. Such coated surfaces have many important applications in colloidal and interfacial science. The polymer layer can change the hydrophobicity of the surface, prevent absorption of other molecules from solution and, in general, plays an important role in colloidal suspensions by preventing flocculation and aggregation of coated colloidal particles \cite{colloidal_domain,fleer_book}.

A densely grafted polymer layer is called a {\it polymer brush}. The layer is grafted
irreversibly on a solid surface by an end--group. Both neutral and charged polymer brushes
have been studied extensively in the last few decades
\cite{NA_phys_rep,SA,PGG,parabolic,parabolic1,pincus,zhulina1,zhulina2,marcelja}. If there
is no strong interaction between the monomers and the surface, the brush properties are
mainly determined by the chain entropy. Neutral brushes, to a large extent, are
characterized by their height that scales linearly with $N$, the polymerization index
\cite{SA,PGG,parabolic,parabolic1}. Charged polymer brushes depend in addition on the
charge density of the chain as well as the solution ionic strength
\cite{pincus,zhulina1,zhulina2,marcelja,szleifer1,szleifer2}.

In this Letter we aim at understanding another variant of  polymer brushes having a terminal charge group, $Ze$, at their free end, where $e$ is the electronic charge and $Z$ the valency (see Fig.~1). The main advantage of having a charged end-group is that we can control the layer height and other properties by applying an external electric field and varying it continuously. This field stretches (or compresses) the chains and is in direct competition with their elastic energy and entropy.
Even in the absence of any external field, we expect the brush to be affected by the charge end-groups, because of their repulsive interactions. Indeed, the height profile depends on the charged group and  an instability of the flat brush towards an undulating one may occur.
\bigskip\bigskip

\section{Flat end-charged polymer brush}
%
\begin{figure}[h!]
\includegraphics[scale=0.45,bb=30 440 565 700,clip]{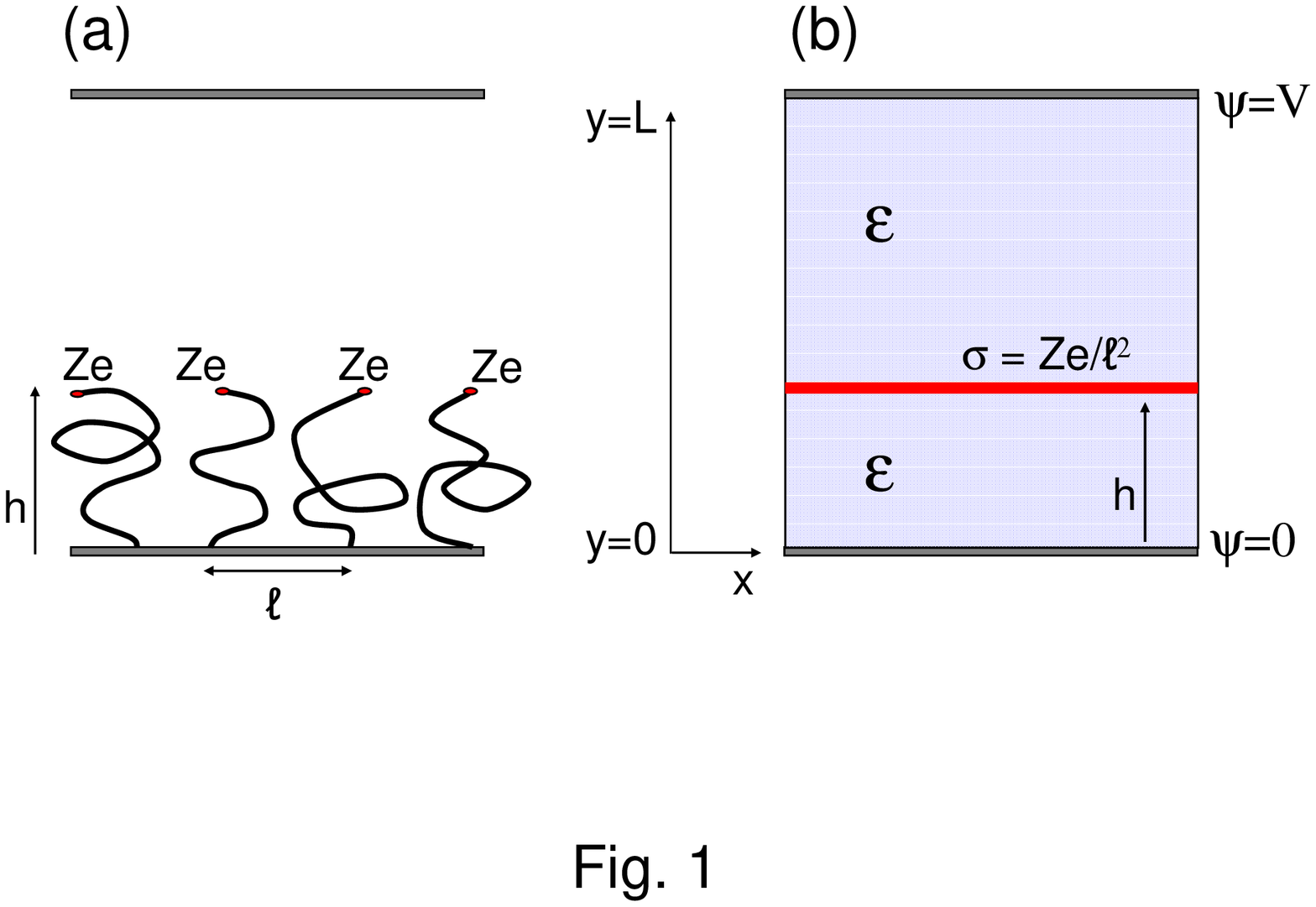}
\caption{(a) Schematic illustration of a polymer brush with a terminal charge group. The polymer is grafted onto a flat and conducting surface at $y=0$, while the other electrode at $y=L$ provides an electrostatic potential difference $V$, with an average electric field, $E=-V/L$. Each chain end-group carries a charge $Ze$, the grafting chain density is $\ell^{-2}$ and the average brush height is $h$. (b) A continuum model of the brush used to calculate the electrostatic properties. The dielectric constant $\eps$ has the same value throughout the gap between the two electrodes. The chain end--charges are bound to a two dimensional layer, with a charge density per unit area, $\sigma=Ze/\ell^2$.
}
\end{figure}

Let us briefly recall the equilibrium
properties  of a {\it neutral} grafted layer. The condition of highly dense layers (the
brush regime) is $\ell \ll R_g$, where $R_g$ is the chain radius of gyration, and $\ell$
is the average distance between chains (Fig.~1a). The brush height, defined as the average
distance of chain ends from the substrate, is denoted by $h$. In the 1970s, a simple
free-energy was proposed by Alexander and de Gennes \cite{SA,PGG} to determine the layer
equilibrium height. In the Alexander -- de Gennes model, the brush height is taken to be
the same for all chains; namely, the height distribution is step-wise. Later, in more
refined theories, it was found that the free-end distribution is parabolic
\cite{parabolic,parabolic1}. In the present work we remain within the step-wise
distribution approximation, which is adequate enough as long as the wavelength of the
predicted instability (see discussion below) is larger than the width of the chain-end
distribution \cite{parabolic,parabolic1}.

The Alexander--de Gennes expression for the brush free-energy is:
\begin{eqnarray}
F_{\rm
brush}=\frac16 Kh^2+\frac{k_BT}{2} v_0 \ell^{-2} N^2 h^{-1} \label{alex}
\end{eqnarray}
where the entropic ``spring'' constant is $K=9k_BT/(Na^2)$, $a$ is the Kuhn statistical length, $N$ the polymerization index, $k_BT$ the thermal energy, and $v_0=a^3(1-2\chi)$ is the excluded volume parameter depending on the Flory-Huggins parameter $\chi$. For a neutral brush, minimization of $F_{\rm brush}$ with respect to the profile height $h$ gives the well-known Alexander--de Gennes height of the brush  at equilibrium $h_{0}$. It scales linearly with $N$:
\begin{equation}
h_{0}=N\left(\frac{1}{6}v_0 a^2 \ell^{-2}\right)^{1/3}
\end{equation}

Next, the end-charged brush is considered. As before, the chains are grafted onto the surface located at $y=0$, but the surface is taken as conducting and is held at potential $\psi=0$ (see Fig.~1) \cite{comment}. At the other (free) end, the chains carry a charge of $Ze$. Without loss of generality we will take this charge to be positive, $Ze>0$. A second conducting and flat surface at $y=L$ is held at a potential $\psi=V$ with respect to the surface at $y=0$. Hence, the polymer brush is subject to a vertical average electric field $E=-V/L$. For $V>0$, the field is compressing the chains, while for $V<0$, it is stretching them.

Although the chain elastic deformation is considered explicitly, the electrostatic properties are calculated within a continuum model where the chains are coarse grained in the following way. The entire gap $0<y<L$ between the two electrodes is assumed to contain the same dielectric medium with dielectric constant $\eps$. The discrete end-group charges are replaced by a two-dimensional layer having a continuous surface charge density. We assume that the continuum limit is adequate enough and provides a good description of the system electrostatics. Finally, we treat the charged brush without taking into account the presence of counter-ions. This assumption and the possible influence of counter-ions is discussed further below for a few relevant limits.

The electric field has a jump at $y=h$, $\Delta E(h)\simeq\sigma/\eps$, where $\sigma=Ze/\ell^{2}$ is the layer charge density per unit area. For a typical grafting density of $\ell \simeq 100$\,nm, dielectric constant $\eps\simeq 10\eps_0$  with $\eps_0$ being the vacuum permittivity and $Z\simeq 5{-}10$, the electric field jump is of order $\Delta E \simeq 10^6\,{\rm V/m}=1\,{\rm V/\mu m}$. Because the charged layer at $y=h$ creates a discontinuity in the electric field $E(y)$, the electrostatic problem is solved separately in two regions: the potential is marked as $\psi_a$ in the region below the charged layer, $0<y<h$, and as $\psi_b$ for the region above it, $h<y<L$. Solving the Laplace equation in the gap $0<y<L$ we get
\begin{eqnarray}
\psi_a&=&by~~~~~~~~~~~~~~~~0<y<h\nonumber\\
\psi_b&=&c~+~dy~~~~~~~~~h<y<L
\label{def_psi}
\end{eqnarray}
where the coefficients $b$, $c$, and $d$ are determined from the boundary conditions: $\psi_a|_{y{=}0}=0$, and $\psi_b|_{y{=}L}=V$. At the charged layer itself $y{=}h$, the potential is continuous, $\psi_a|_{y{=}h}=\psi_b|_{y{=}h}$, and the jump in its electric field is proportional to the charge density $\sigma$:
\begin{eqnarray}\label{coeffs_bcd}
b=\frac{V}{L}+\frac{\sigma}{\eps}(1-h/L)~~;~c=\frac{\sigma h}{\eps}~~;~
d=\frac{V}{L}-\frac{\sigma h}{\eps L}
\end{eqnarray}

The total free-energy can be written as the sum \cite{LL}
\begin{eqnarray}
F=F_{\rm brush}-\frac12\int
\eps\left(\nabla\psi\right)^2~d^3r+\int\rho\psi~d^3r
\end{eqnarray}
where $F_{\rm brush}$ is the brush free energy, and the last two terms represent the
electrostatic energy (in SI units). The volume charge density $\rho$ is related to the surface one via the Dirac delta--function, $\rho=\sigma\delta(y-h)$.

We first consider the case where the electrostatic interactions have a small effect on the thickness and we expand $F_{\rm brush}$ around its value at $h_{0}$ to second order in $h-h_{0}$. The resulting total free-energy per grafting site is
\begin{eqnarray}
F&=&\frac12 K(h-h_{0})^2~-~\frac12
\ell^2\eps\frac{V^2}{L}\nonumber\\
&+&\ell^2\left(\frac{\sigma
V}{L}+\frac12\frac{\sigma^2}{\eps}\right)h
~-~\frac12\frac{\ell^2
\sigma^2}{\eps L}h^2~+~const.\label{FE}
\end{eqnarray}
Minimization of $F$ with respect to the brush height $h$ yields the equilibrium
brush height  $h_{\rm el}$ for the charged case:
\begin{equation}\label{h0_approx}
h_{\rm el}\simeq h_{0}\left(
1-\frac{\sigma^2\ell^2}{K\eps}\left[\frac{1}{2h_0}-\frac{1}{L}
\right]
-\frac{\sigma \ell^2V}{KLh_{0}} \right)
\end{equation}
This expression is valid for low enough $\sigma$, i.e. when $\sigma^2\ll K\eps
h_0/\ell^2$ and $\sigma V\ll KLh_0/\ell^2$.

Note that the brush height is compressed, $h_{\rm el} <h_{0}$, even when the external
potential gap between the two electrodes vanishes, $V=0$. In this case the charges at the
brush end--groups are attracted to the induced image charges on the two
grounded electrodes, and the interaction with the closer electrode (at the origin) is
stronger.

In the opposite case of strong charge-charge interactions, the brush height can be
much smaller than $h_{0}$, which makes eq~(\ref{FE}) invalid. In this case the brush free
energy is dominated by excluded volume interactions [last term in eq~(\ref{alex})]. In
the limit $h\ll L$ the brush height is:
\begin{equation}\label{h_compressed}
h_{\rm el}^2=\frac12 k_BT v_0 \ell^{-4} N^2 \frac{1}{\left(\frac{\sigma
V}{L}+\frac12\frac{\sigma^2}{\eps}\right)}
\end{equation}
%

\section{Undulating charged brush}

Because neighboring chains carry identical charges they repel each other. The
electrostatic energy can be further reduced if the chains compress or stretch
alternatively, thereby increasing the distance between chain ends. In order to investigate
whether such an interfacial instability exists, we assume that the brush height has a
single undulation mode along one of the lateral surface dimensions, $x$: 
\begin{eqnarray}\label{h_of_x}
h(x)=h_{\rm el}+h_q\cos(qx)
\end{eqnarray} 
where $h_{\rm el}$ is the equilibrium location of a flat brush [eq.
(\ref{h0_approx})], $q=2\pi/\lambda$ is the modulation $q$-mode, and $h_q$ the amplitude.
As was done above for the flat layer, the gap is divided
 into two regions and the potential is $\psi=\psi_a$ for $y<h(x)$ and $\psi=\psi_b$ for
$h(x)<y<L$. The potential $\psi(x,y)$ satisfies Laplace's equation $\nabla^2\psi=0$, with
the following four boundary conditions:
\begin{eqnarray}\label{bcs}
\left. \psi_a\right|_{y=0}=0 ~~;~~ \left.\psi_b\right|_{y=L}=V\nonumber\\
\left.\psi_a=\psi_b\right|_{y=h(x)} \quad,\nonumber\\ \left.
{\eps\hat{n}\cdot\nabla(\psi_a\,-\,\psi_b)}\right|_{y=h(x)}=\sigma_{\rm eff}(x)\quad. \end{eqnarray}
\begin{figure}[h!]
\includegraphics[scale=0.7,bb=145 440 435 650,clip]{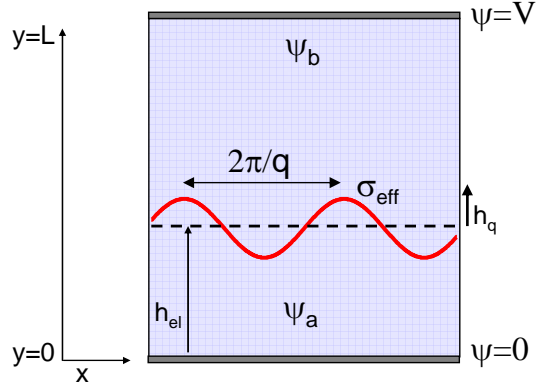}
\caption{A brush with an undulating height profile given
 by $h(x)=h_{\rm el}+h_q\cos qx$ confined between
two conducting and flat surfaces located at $y=0$ and $y=L$.
}
\end{figure}
Here $\hat{n}=(qh_q\sin qx,1)/\sqrt{1+(qh_q\sin qx)^2}$ is a unit vector normal to the
undulating interface given by $h(x)$. The density $\sigma_{\rm eff}(x)$ appearing in
the above boundary condition is defined as: $\sigma_{\rm eff}=\sigma/\sqrt{1+(qh_q\sin
qx)^2}$ and is related to the constant density $\sigma$ on the projected area, as defined
in the previous section for the flat layer. In our simplified treatment, the media
dielectric constant $\eps$ has the same value below and above the brush surface. Even
within  the uniform dielectric media assumption, we are able to show that the flat
interface can be unstable. This demonstrates that the instability, due to charge-charge
interactions, is different from other instabilities considered in
Refs.~\cite{steiner1,steiner2,steiner3}, and related to heterogeneous (uncharged)
dielectric materials placed in external electric fields.

The potential within the polymer layer, $\psi_a$, and above it, $\psi_b$, are written as a power series in $h_q$: $\psi_a=\sum_{n=0}^{\infty}\psi_a^{(n)}(h_q)^n$, and $\psi_b=\sum_{n=0}^{\infty}\psi_b^{(n)}(h_q)^n$. Clearly the Laplace equation is satisfied separately for each order $n$ in the expansion: $\nabla^2\psi_a^{(n)}=0$ and $\nabla^2\psi_b^{(n)}=0$.
Note also that the zeroth order terms, $\psi_{a}^{(0)}$ and $\psi_{b}^{(0)}$, are the
solution of the flat charged layer [eqs. (\ref{def_psi}) and (\ref{coeffs_bcd})]. It then
follows that for $n>0$
\begin{eqnarray}
\psi_a^{(n)}=\sum_{k\neq 0}\left(a_k^{(n)}\exp(ky)+b_k^{(n)}\exp(-ky)\right)\cos
kx\nonumber\\
\psi_b^{(n)}=\sum_{k\neq 0}\left(d_k^{(n)}\exp(ky)+e_k^{(n)}\exp(-ky)\right)\cos kx
\end{eqnarray}

The leading contributions in $h_q$, the layer undulation amplitude, can be examined by assuming that $h_q\ll h_{\rm el}$ and expanding the electrostatic free energy up to order $\sim(h_q)^2$. We, therefore, limit ourselves to the first order in $h_q$: $\psi=\psi^{(0)}+ \psi^{(1)}h_q$. Furthermore, we focus on the long wavelength limit, $qh_q\ll 1$, relevant to small amplitude modulations.

For linear order in $h_q$, only the first Fourier component $k=q$ does not vanish,
and we find
\begin{eqnarray}
a_q^{(1)}&=&-b_q^{(1)}=-\frac{\sigma}{2\eps}\frac{\cosh q(L-h_{\rm el})}{\sinh
qL}\nonumber\\
d_q^{(1)}&=&-e_q^{(1)}=-\frac{\sigma}{2\eps}\frac{\cosh qh_{\rm el}}{\sinh qL}\,\,{\rm
e}^{-qL}
\end{eqnarray}
Expanding to second order in $h_q$, both the  $k=0$ mode and the second harmonics $k=2q$
do not vanish.

The electrostatic energy difference $\Delta f_{\rm el}$ (per unit area) between the undulating
($h_q\neq 0$) and the flat layer ($h_q=0$) is given to second order in $h_q$ by
\begin{eqnarray}
&&L_xL_z\Delta f_{\rm el}=\nonumber\\
&-&\frac{\eps}{2}\int\int{\rm d}x\,{\rm
d}z\int_{y=0}^{y=h}{\rm
d}y\,\left[2 h_q\nabla\psi_a^{(0)}\cdot\nabla\psi_a^{(1)}\right.\nonumber\\
&&\left.+h_q^2\left(\nabla\psi_a^{(1)}\right)^2
\right]\nonumber\\
&-&\frac{\eps}{2}\int\int{\rm d}x{\rm
d}z\int_{y=h}^{y=L}\,{\rm d}y\left[ 2h_q\nabla\psi_b^{(0)}\cdot\nabla\psi_b^{(1)}\right.\nonumber\\
&& \left.+h_q^2\left(\nabla\psi_b^{(1)}\right)^2
\right] \nonumber\\
&+&\sigma h_q\int{\rm d}x\int{\rm
d}z\left[a_q^{(1)}\exp(qh)+b_q^{(1)}\exp(-qh)\right]\cos qx~\nonumber\\
\end{eqnarray}
where $L_x$ and $L_z$ are the two lateral dimensions and $h$ is $h(x)$ from eq.
(\ref{h_of_x}).
Straightforward algebraic manipulations give the final answer for the electrostatic energy
difference per unit area of the brush in the long wavelength limit ($qh_q\ll 1$):
\begin{eqnarray}\label{DF_es}
\Delta f_{\rm el}=-\frac{\sigma^2}{\eps}\frac{\cosh(qh_{\rm el})\cosh[q(L-h_{\rm
el})]}{\sinh(qL)}qh_q^2
\end{eqnarray}
The scaling of the last expression could have been guessed from the outset. The electrostatic energy is symmetric in $h_q\to -h_q$ and, to lowest orders, is quadratic in $h_q$. In addition, the prefactor $\sigma^2/\eps$ has dimensions of dielectric constant times electric field squared, and thus $\Delta f_{\rm el}$ must be linear in $q$. The derivation above gives us in addition the numerical factors and an extra dependence containing trigonometric functions. These are symmetric with respect to the transformation $h_{\rm el}\to L-h_{\rm el}$.

Note that the externally imposed potential $V$ does not appear explicitly in $\Delta f_{\rm el}$. The external field simply stretches the brush uniformly, thereby increasing $h_{\rm el}$. The interfacial instability is solely due to charge-charge interactions, as exemplified by the $\sigma^2$ prefactor. The simple case of a thin isolated charged layer embedded in an infinite medium of uniform dielectric constant is obtained in the symmetric limit $h_{\rm el}=\frac12 L$, and $L\to\infty$. In this case one finds $\Delta f_{\rm el}=-(\sigma^2/2\eps)\cdot qh_q^2$.

\section{Brush surface instability}
The brush instability mentioned above causes a deformation of the flat layer and costs
interfacial and elastic energy. We consider first the effect of surface tension and then
comment on the elasticity. For a single $q$ mode, the interfacial energy per unit area is
$\Delta f_\gamma=\frac{1}{4}\gamma q^2h_q^2$. The total free energy difference $\Delta
f=\Delta f_\gamma+\Delta f_{\rm el}$ is:
\begin{eqnarray}\label{DF}
\Delta f/h_q^2\simeq
-\frac{\sigma^2}{\eps }\frac{\cosh(qh_{\rm el})
\cosh[q(L-h_{\rm el})]}{\sinh(qL)}q+\frac{1}{4}\gamma q^2
\end{eqnarray}
$\Delta f$ has a minimum at a finite wavenumber $q^*$ as can be seen in Fig.~3(a), where
$\Delta f$ is plotted with a choice of typical parameter values. The value of $q^*$ can be
obtained by solving the transcendental equation
\begin{equation}
q^*=\frac{2}{\gamma h_q^2}\left.\frac{\partial \Delta f_{\rm el}}{\partial
q}\right|_{q^*}
\end{equation}
In Fig.~3(b) we show the dependence of $q^*$ on $h_{\rm el}/L$. The most stable wavenumber
$q^*$ decreases monotonically as $h_{\rm el}$ increases.

\begin{figure}[h!]
\includegraphics[scale=0.68,bb=125 215 450 620,clip]{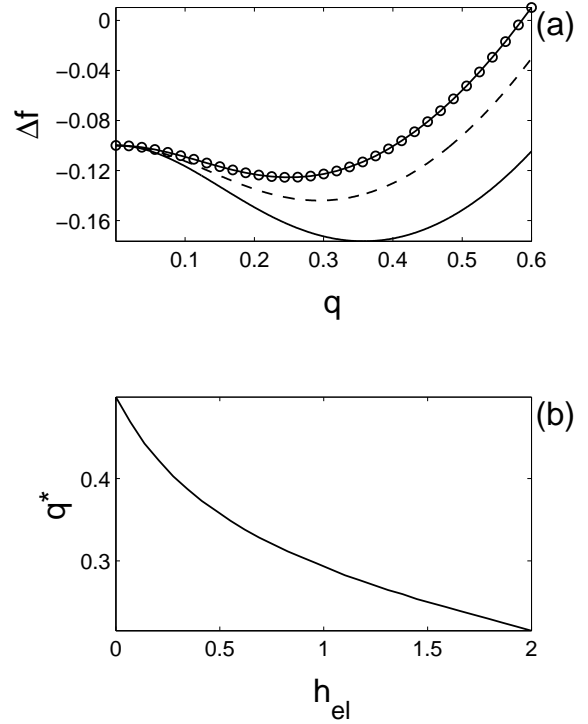}
\caption{(a) The  brush free-energy $\Delta {f}$ from eq.~(\ref{DF}) in dimensionless
units
[rescaled by $(\sigma^2/\eps)^2(h_q^2/\gamma)$],
as a function of the dimensionless wavenumber $q$ [rescaled by $\sigma^2/\eps\gamma$]. Solid, dashed and circle-decorated curves correspond to rescaled brush height $h_{\rm el}(\sigma^2/\eps\gamma)=0.5$, $1$ and $1.5$, respectively, and rescaled film thickness $L(\sigma^2/\eps\gamma)=10$. The location of the minimum increases as $h_{\rm el}$ decreases. (b) The most stable dimensionless wavenumber ${q}^*$ as a function of rescaled height $h_{\rm el}$.
}
\end{figure}

In order to check whether the predicted instability can be seen in experiments, we estimate its order of magnitude by taking the brush charge to be $Z=5$, chain separation $\ell=20$\,nm, dielectric constant $\eps=5\eps_0$ and surface tension of the brush layer $\gamma=1$\,mN/m. We use $h_{\rm el}=\gamma\eps/\sigma^2\simeq 10$\,nm and $L=10h_{\rm el}$, and find $q^*\simeq 2.8\cdot 10^6$\,m$^{-1}$, resulting in an undulation wavelength $\lambda^*=2\pi/q^*\simeq 2.2$\,$\mu$m. This is indeed the long wavelength limit which agrees with the various approximations we made. By further changing the system parameters: $\ell$, $Ze$ and $\gamma$ it is possible to tune $q^*$ and adapt its value (in the micrometer range) in specific experimental setups.

\section{Discussion and Conclusions}

In this Letter we revisit the problem of polymer brushes. The new feature considered was
to attach a charge $Ze$ to the chain terminal free-end. The brush can be grafted onto an electrode (flat surface), while a second distant electrode is placed above the brush and provides a voltage gap $V$. The net effect is to have a controlled way to compress or expand the layer height, $h$.

The charged brush creates an effective surface charge density that is localized at the
$y{=}h$ interface. These charges repel each other and also interact with the external
electric field. The equilibrium height $h_{\rm el}$ depends on the competition
between all electrostatic interactions and the elasticity and entropy of the neutral
chain, as can be seen from eqs.~(\ref{h0_approx}) or (\ref{h_compressed}). It can lead to
a compression or expansion of the layer height with respect to the neutral brush case. But
maybe the more interesting effect is the onset of an instability in the layer height.
Employing a linear stability analysis, the conditions leading to an instability of the
uniform layer are analyzed, and a preferred wavenumber $q^*$ stabilized by surface
tension is found. When elasticity of the polymer brush is included, it
contributes  a $q^4$ term in eq. (\ref{DF}) \cite{LL_elasticity} in addition to the $q^2$
term originating from surface tension. The qualitative system behavior is similar, with a
modified expression for the preferred wavenumber $q^*$.

The full system behavior in the presence of counterions is quite complicated and should be
explored in a separate work, especially the limit of highly charged brushes.  Here, we 
comment briefly on two extreme limits. In the first limit, the brush charge and the
external potential are taken to be small enough so that the counterions are uniformly
distributed throughout the available volume, and behave like an  ideal gas. It then
follows that the electrostatic potential depends quadratically on the direction $y$. In
this approximation, the counterions do not contribute to the pressure difference
$\Delta P$ across the brush end. The only source of this pressure difference is
electrostatic and is due to the difference in $\frac{1}{2}\eps E^2$ between the two sides
of the brush. We find $\Delta P=\sigma V/L+\sigma^2/\eps-3\sigma^2h/\eps L$.

In a second scenario, the brush charge is small, but the electrostatic energy of
counterions in contact with the electrode, $eV$, is much larger than the thermal energy
$k_BT$. Here we find that all counterions migrate to one of the electrodes. However, since
the above calculation assumes a fixed voltage gap, $V$, the electrodes will accumulate
extra charge to balance exactly the counterions. Therefore, the results in eqs.
(\ref{FE}), (\ref{h0_approx}) and (\ref{h_compressed}) stay valid. Lastly, we point out
that in a more physically feasible setup, the system may contain added salt
\cite{szleifer1,szleifer2}. In this case, the electric field is screened and the brush
ends do not ``feel'' the electrodes as long as the brush length is larger than the
Debye-H\"{u}ckel screening length.

It is worthwhile to mention some similarities between our charged brush and other two-dimensional systems of charges or dipoles. A two-dimensional layer of electric dipoles  pointing in the perpendicular direction was investigated~\cite{JCP87} in relation with dipolar Langmuir monolayers at the water/air interface \cite{JCP87}. When the dipoles have a fixed out-of-plane moment but their inplane density can vary, a modulated phase in the inplane density can be stabilized with a preferred wavenumber $q^*$. In addition, the dipolar free energy also scales linearly in $q$~\cite{JCP87}. The similarity between the two systems can be understood in the following way. The charge displacement from their average position at $y=h_{\rm el}$ in our case is similar to an effective dipole whose moment points `up' or `down' with respect to this reference plane. Increase in the external $E$ field in our system translates into an increase proportional to the average dipole strength in the dipolar system.

More recently~\cite{safran07}, a $q$-mode instability was found for an electric double layer where a charge density bound to a surface was allowed to fluctuate laterally. The model is motivated by an experimental setup where charged amphiphiles coat heterogeneously a mica surface. In the experiment the surface contains patches of positive and negative charges but the overall surface charge (summed over all patches) remains zero. As the surface was placed in contact with a salt solution, a local electric double layer is formed. Positively and negatively charged counterions are attracted, respectively, to negative and positive charge domains, resembling our system as well as the undulating dipolar one~\cite{JCP87,safran07}.

We hope that the simple considerations mentioned here will motivate experimental studies of end-charged brushes where some of our predictions can be tested.

\acknowledgments
We would like to thank Terry Cosgrove for suggesting us this problem and Dan Ben-Yaakov for helpful discussions. YT acknowledges support from the Israel Science Foundation (ISF) under grant no. 284/05, and German-Israeli Foundation (GIF) grant no. 2144-1636.10/2006. DA acknowledges support from the Israel Science Foundation (ISF) under grant no. 160/05 and the US-Israel Binational Foundation (BSF) under grant no. 2006055.


\begin{thebibliography}{0}
\bibitem{comment} The presence of a thin insulating layer to which the chains are
grafted is not expected to change our results.
\end{thebibliography}
\end{document}